%% file: main.tex
\documentclass[11pt,aps,tightenlines,nofootinbib,longbibliography,superscriptaddress]{revtex4-1}

\usepackage[utf8]{inputenc}
\usepackage{charter} 
\usepackage{fullpage}
\usepackage{hyperref}
\usepackage{graphicx}
\usepackage{lipsum}
\usepackage{hyperref}
\hypersetup{hidelinks}
\usepackage[total={6.5in,9in},top=1in,headsep=0.1in,headheight=1in]{geometry}
\usepackage{enumitem,amssymb}
\usepackage[usenames,dvipsnames]{xcolor}
\usepackage{aas_macros}
\usepackage{xspace}
\usepackage{multirow}
\usepackage[normalem]{ulem}

\newcommand\snowmass{
\begin{center}
  \rule[-0.2in]{\hsize}{0.01in}\\
  \rule{\hsize}{0.01in}\\
  \vskip 0.1in
  Submitted to the Proceedings of the US Community Study\\ 
  on the Future of Particle Physics (Snowmass 2021)\\
  \rule{\hsize}{0.01in}\\
  \rule[+0.2in]{\hsize}{0.01in}\\[-2em]
\end{center}
}

\usepackage[firstpage=true]{background}
\backgroundsetup{contents={\parbox{6.5in}{\snowmass}}, scale=1,placement=top,opacity=1,color=black,position={3.25in,1.2in}}

\usepackage{fancyhdr}
\fancypagestyle{plain}{%
  \fancyhf{}%
  \fancyhead[C]{}
  \fancyfoot[C]{\thepage}
}

\fancypagestyle{empty}{%
  \fancyhf{}%
  \fancyhead[C]{{\it Snowmass2021: Opportunities from Cross-survey Analyses of Static Probes}}
  \fancyfoot[C]{\thepage}
}
\usepackage[utf8]{inputenc}
\usepackage{newunicodechar,graphicx}
\DeclareRobustCommand{\okina}{%
  \raisebox{\dimexpr\fontcharht\font`A-\height}{%
    \scalebox{0.8}{`}%
  }%
}
\newunicodechar{ʻ}{\okina}
\newcommand*\hawaii{Hawai\okina{}i}

\newcommand{\inlinemaketitle}{{\let\newpage\relax\maketitle}}

\begin{document}

\title{Snowmass2021 Cosmic Frontier White Paper:\\ Opportunities from Cross-survey Analyses of Static Probes}

\input{authors.tex}

\begin{abstract}
Cosmological data in the next decade will be characterized by high-precision, multi-wavelength measurements of thousands of square degrees of the same patches of sky.  By performing multi-survey analyses that harness the correlated nature of these datasets, 
we will gain access to new science, and increase the precision and robustness of science being pursued by each individual survey.  However, effective application of such analyses requires a qualitatively new level of investment in cross-survey infrastructure, including simulations, associated modeling, coordination of data sharing, and survey strategy.  The scientific gains from this new level of investment are multiplicative, as the benefits can be reaped by even present-day instruments, and can be applied to new instruments as they come online.
\end{abstract}

\inlinemaketitle


\section{Introduction}

The next decade will see a dramatic improvement in our ability to probe the Universe, with major leaps in capabilities occurring nearly simultaneously across many new facilities.  Each of these new facilities will enable transformative science, but joint analyses of the resultant datasets will be more powerful and robust than what can be achieved with any individual instrument.  In this whitepaper we focus on the case for, and implications of, joint analyses and cross-correlations between datasets that span different experiments and projects within the Cosmic Frontier.  The promise of such analyses is both that they enable new science, as well as  increase the robustness of the core science being pursed by each project.  Notably, cross-survey analyses will improve the constraints on cosmic acceleration that drive the design and requirements for cosmological surveys into which DOE has invested, and also leverage those investments to constrain other aspects of fundamental physics that are  important for our understanding of the Universe.  At present, however, cross-survey analyses can be challenging to initiate, organize and fund.  One of the main goals of this whitepaper is to advocate for the creation of clear pathways to support cross-survey analyses as part of the core mission of DOE's Cosmic Frontier. 

\begin{figure*}
    \centering
    \includegraphics[scale=0.65]{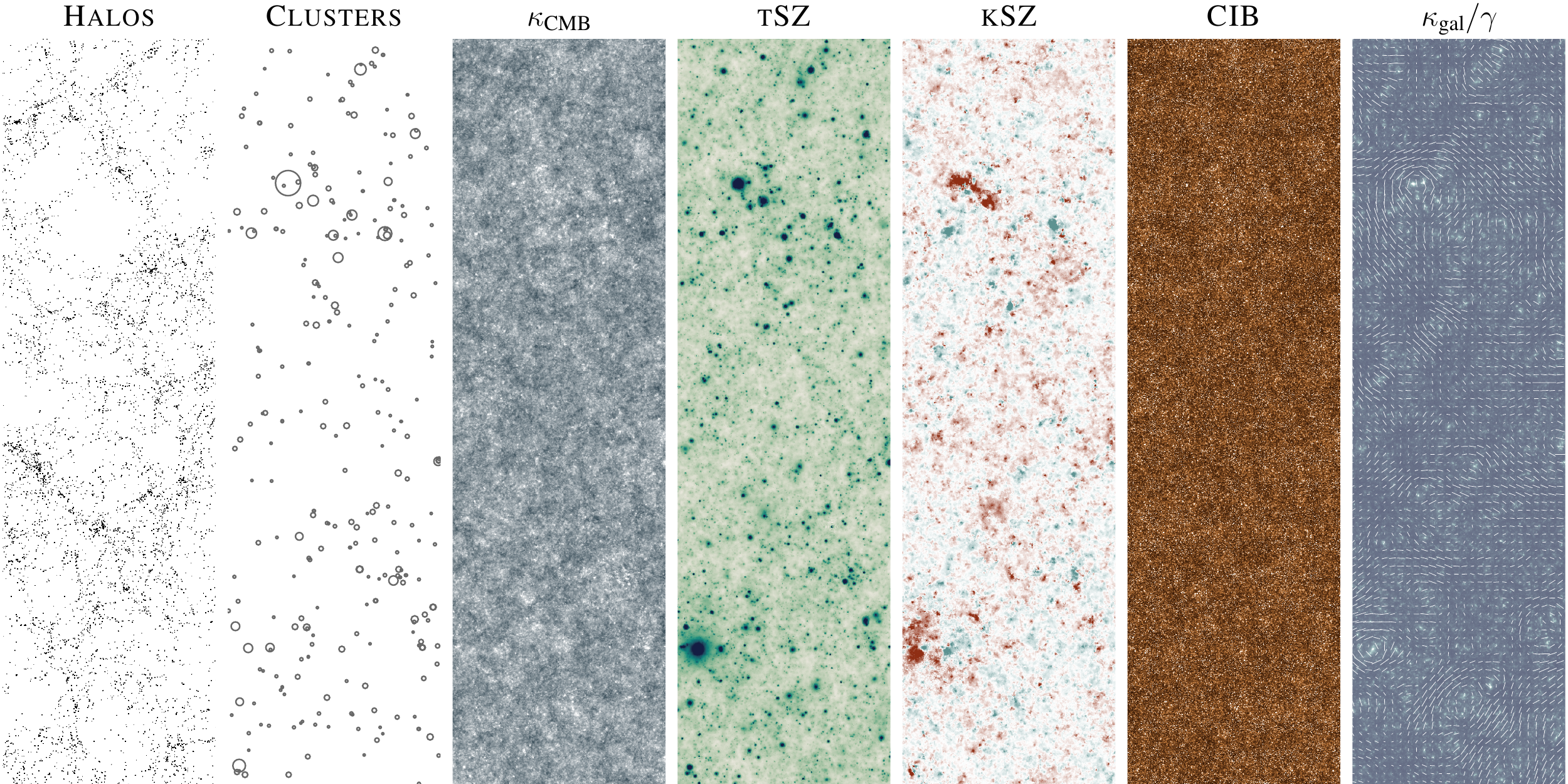}
    \caption{Simulated maps of the same patch of the Universe, as measured with several different cosmological probes (from left to right): dark matter halos (detectable via the galaxies they host), galaxy clusters (with the size of the circles indicating the cluster mass), gravitational lensing of the CMB ($\kappa_{\rm CMB}$), the thermal Sunyaev Zel'dovich effect (tSZ), the kinematic Sunyaev Zel'dovich effect (kSZ), the cosmic infrared background (CIB), and gravitational lensing of galaxy shapes (shading indicates the convergence, $\kappa_{\rm gal}$, while white lines indicate the shear, $\gamma$).  Although each probe is very different, they are all sourced by the same underlying large scale structure, and are therefore correlated.  Joint analyses of these different probes can yield access to new cosmological information about the underlying structure. Simulated data from Omori (in prep.). }
    \label{fig:multiwavelength}
\end{figure*}

As an illustration of the diversity of possible cross-survey analyses, Fig.~\ref{fig:multiwavelength} presents simulated maps of a patch of the Universe as measured by different cosmological probes.  Each probe is connected to the same underlying large scale structure, and as a result, all of these probes are correlated.  By cross-correlating probes from different surveys, new information about cosmological structure can be extracted. The improvements to our understanding of the Universe that are enabled by joint analyses of multiple surveys are remarkably diverse.  Some prominent examples include: 
\begin{itemize}
    \item {\bf Improved robustness of cosmological constraints}.  Analyses of cross-correlations between surveys will increase the robustness of cosmological constraints by breaking degeneracies with nuisance parameters that degrade single-survey constraints.  
    Additionally, cross-correlations can provide tight constraints on astrophysical sources of systematic uncertainties, such as baryonic feedback and intrinsic alignments of galaxies. 
    \item {\bf Improved cosmological constraints from the evolution of large scale structure}.  Cross-correlations of galaxy surveys with gravitational lensing of the CMB offer the prospect of tight constraints on structure at high redshift, improving constraints on dark energy, modified gravity, and the sum of the neutrino masses. Cross-correlations with line intensity mapping experiments offer similar benefits if long wavelength modes along the line of sight can be recovered.
    \item {\bf Improved cosmological constraints from the abundance and clustering of galaxy clusters}.  Galaxy clusters have the potential to be powerful cosmological probes, but realizing this potential will require control of astrophysical and systematic uncertainties.  Cross-survey, multi-wavelength studies of galaxy clusters offer the prospect of significantly improved constraints on cluster masses and other properties.
    \item {\bf Improved cosmological constraints from overlapping imaging and spectroscopic surveys.}   Spectroscopic surveys can provide high-accuracy redshift information and improved source classification for objects detected in overlapping imaging surveys, enabling improved cosmological constraints.  At the same time, imaging surveys provide a complete census of galaxies that allows for compilation of targets for spectroscopic surveys, as well as measurements of structure inaccessible to spectroscopic surveys. 
    \item {\bf Improved constraints on non-Gaussianity}.  By exploiting multi-tracer techniques and the high-redshift reach of CMB lensing, cross-correlations offer the prospect of tight constraints on primordial non-Gaussianity and inflationary models.
    \item {\bf A census of baryons}.  The thermal and kinematic Sunyaev Zel'dovich effects measured by CMB surveys provide a 2D snapshot of the distribution and thermal state of baryons throughout the Universe.  By cross-correlating these measurements with probes of known redshift --- such as galaxies --- 3D information about the baryons can be recovered.  Upcoming surveys will also allow for a first detection of the polarized SZ effect, providing a new tool to study the distribution of baryons.  
\end{itemize}

\begin{figure}
\centering
\includegraphics[width=15cm]{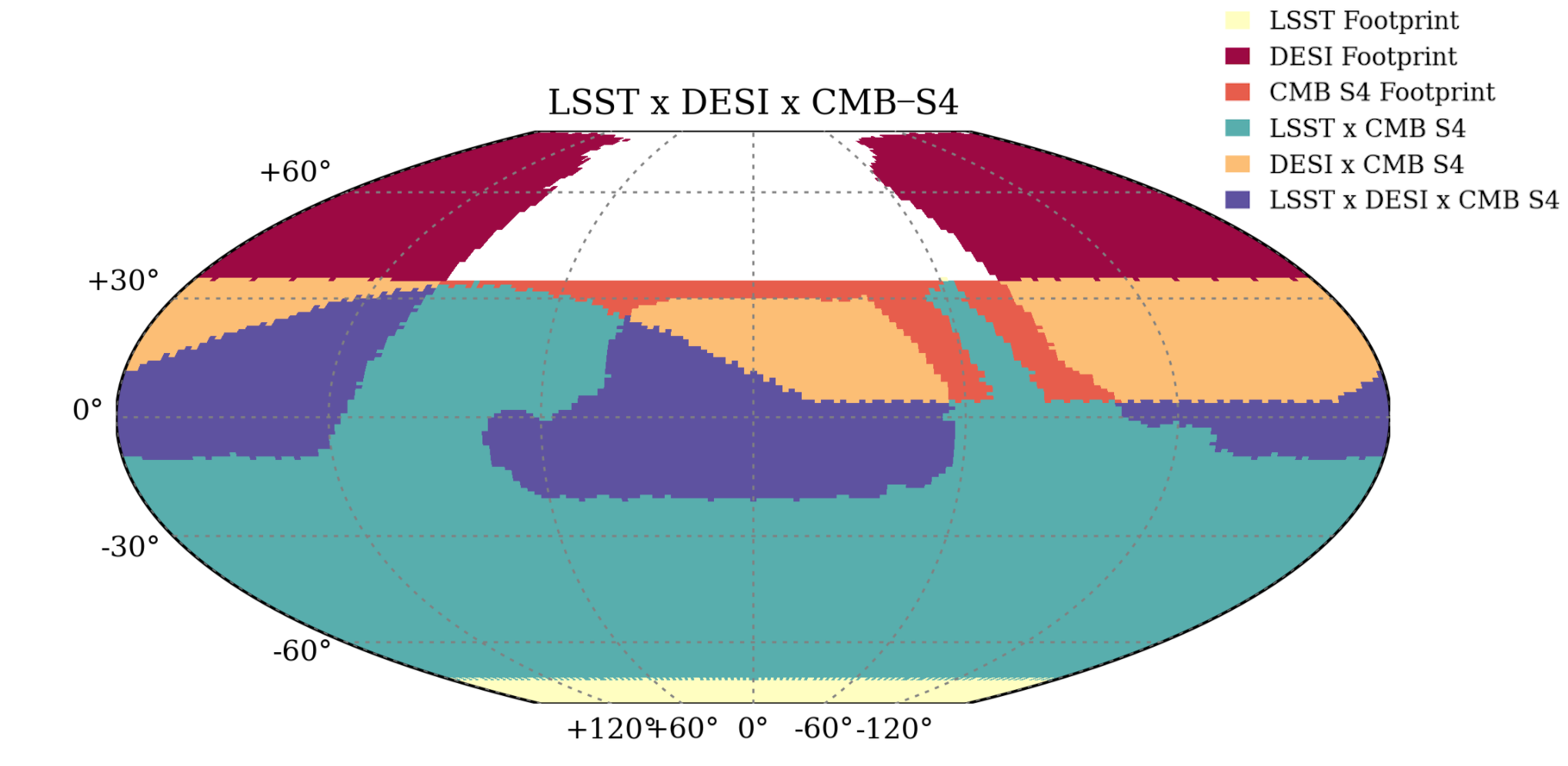}
\caption{
Future cosmic surveys will have large regions of overlapping coverage, enabling opportunities for cross-correlations.  Here, we illustrate this overlap for three prominent future optical imaging (LSST), spectroscopic (DESI), and CMB (CMB-S4) surveys.  The other future surveys listed in Table~\ref{tab: survey_summary}, omitted for clarity, are also expected to have significant overlap.   
}
\label{fig:footprint2}
\end{figure}

Measuring cross-correlations between different cosmological probes requires overlapping measurements on the sky.  As shown in Fig.~\ref{fig:footprint2}, the survey strategies of several operational and planned DOE-funded cosmic surveys  --- including optical imaging, spectroscopic, and CMB surveys --- have significant overlap.  While we illustrate the overlap for three specific surveys, many of the opportunities and challenges discussed here can be applied to any cross-survey analysis; we list other relevant surveys (and the corresponding acronyms used throughout the text) in Table~\ref{tab: survey_summary}. Given the significant overlap on the sky of future cosmic surveys, there is potential to harness the power of cross-correlations between them.  However, as we discuss below, actually performing such analyses to maximize the science return from cross-correlations will require significant additional investments in simulation and analysis infrastructure, as well as mechanisms for improved cross-survey collaboration.   

\begin{table*}
  \caption{List of recently commenced and planned cosmological surveys.}
  \label{tab: survey_summary}
  \centering
  \begin{tabular}{p{0.3\linewidth}|p{0.5\linewidth}|p{0.1\linewidth}}
\hline 
Type & Experiment (Acronym) & Reference \\
\hline
\multirow{4}{*}{Optical/NIR Imaging} & Vera C. Rubin Observatory's Legacy Survey of Space and Time (LSST)  & \cite{LSST,lsst-science} \\
& Euclid &  \cite{Euclid} \\
& Nancy Grace Roman Space Telescope (Roman) &  \cite{Dore2019} \\
\hline
\multirow{5}{*}{Optical/NIR Spectroscopy} & Dark Energy Spectroscopic Instrument (DESI) & \cite{desi19}  \\
& 4-metere Multi-Object Spectroscopic Telescope (4MOST) &  \cite{4MOST} \\
& Maunakea Spectroscopic Explorer (MSE) &  \cite{MSE} \\
& MegaMapper &  \cite{Schlegel:2019} \\
& Nancy Grace Roman Space Telescope (Roman) &  \cite{Dore2019} \\
\hline
\multirow{3}{*}{ \shortstack[l]{Cosmic Microwave Background \\ (Large Aperture) }} & Simons Observatory (SO)  & \cite{SimonsObs} \\
& CMB-Stage 4 (CMB-S4) &  \cite{S4} \\
&  CMB-High Definition (CMB-HD) &  \cite{CMB-HD-Snowmass} \\
\hline
\multirow{2}{*}{Line Intensity Mapping} & Spectro-Photometer for the History of the Universe, Epoch of Reionization, and Ices Explorer (SPHEREx)  & \cite{SPHEREx} \\
&  Packed Ultra-wideband Mapping Array (PUMA) &  \cite{PUMA} \\
\hline
\multirow{2}{*}{X-ray} & eROSITA  & \cite{EROSITA} \\
&  Athena &  \cite{Athena} \\
\hline
\end{tabular}
\end{table*}

We describe examples of specific cross-survey analyses that enable these many improvements to cosmological constraints in \S\ref{sec:LOIs}.  Extracting cosmological information from these measurements will present new modeling and analysis challenges, which we discuss in \S\ref{sec:model_analysis}.  Not unexpectedly, these joint analyses have associated costs, implications and impacts on the individual projects.  We discuss several growth opportunities for addressing these issues and maximizing  science return in \S\ref{sec:recs}.

\section{Science from joint probes}
\label{sec:LOIs}

\subsection{CMB lensing $\times$ galaxies}

Lensing of the CMB by the gravitational potentials associated with large-scale structure measured in future galaxy surveys offers several key science opportunities.  First, it provides a means of measuring the gravitational potentials, and hence matter density, at high redshifts where this becomes infeasible with galaxy lensing.  This, in turn, gives strong constraints on the amplitude of the power spectrum with redshift \cite{Omori:2019, White:2021yvw, Krolewski:2021yqy} which is crucial to tests of modified gravity \cite{Pullen:2015}, neutrinos \cite{Yu:2018tem} and dark energy \cite{Yu:2021vce}, and directly bears upon the tensions being raised by lower redshift probes.  By using relativistic tracers (photons), CMB lensing provides access to the space-space perturbations to the metric, which in combination with redshift-space distortions, provides a test of gravity.  By giving access to different tracers of the fluctuations at the highest redshifts and largest scales, a combination of galaxy clustering and CMB lensing may provide our tightest constraints on primordial non-Gaussianity \cite{Schmittfull2017,Snowmass2021:Inflation}.  Joint analyses of cross-correlations between galaxy surveys and CMB lensing can also exploit the fact that parameter dependencies of cross-correlations are typically different from those of intra-survey correlations, enabling significant degeneracy breaking \cite{2016MNRAS.461.4099B}.

The technique of CMB lensing tomography \cite{Giannantonio:2016}, enabled by CMB surveys such as CMB-S4 and CMB-HD, and galaxy catalogs from surveys such as LSST, Euclid and Roman, will allow for the creation of mass maps in broad redshift slices out to redshifts as high as 5,  making possible new precision tests of cosmology.  Such results explore the connection between visible baryons and the underlying dark-matter scaffolding.  CMB lensing also enables calibration of cluster masses at high redshift \cite{Baxter:2015}, allowing the abundance of galaxy clusters to be used as an additional  probe of dark energy and neutrino masses (see also \S~\ref{sec:clusters}).

In addition to new science opportunities and improved cosmological constraints, cross-correlations between galaxy surveys and CMB lensing have the potential to make cosmological analyses more robust.
The complex astrophysics that determines where galaxies form (``biasing'') and their intrinsic shapes (``intrinsic alignments'') can substantially affect the cosmological interpretation of galaxy surveys and the cosmological constraints inferred from them at the level of the 2-point correlations \cite{Krause:2016,Yao:2017,Blazek:2019} as well as higher-order statistics  \cite{2012MNRAS.427..442T,2012MNRAS.423.1663T,2012MNRAS.419.1804T}. Correlating galaxy positions and lensing with CMB lensing, an independent mass tracer, provides a powerful method to improve our understanding of these effects and to mitigate systematic biases.  In a similar vein, cross-correlations of galaxy surveys with CMB lensing can be used to calibrate multiplicative biases that impact galaxy lensing measurements, allowing the data to self-calibrate, rather than relying on e.g. image simulations to constrain these biases \cite{Schaan:2017}.  Finally, cross-correlations between galaxy surveys and CMB lensing measurements have the advantage of being largely immune to additive systematics in the galaxy survey or CMB lensing observables.  As long as such systematics do not impact both fields being correlated, they are suppressed in cross-correlation.

\subsection{Thermal/Kinetic Sunyaev Zel'dovich effect $\times$ galaxies}

The Sunyaev Zel'dovich (SZ) effect \cite{SZ} is caused by inverse Compton scattering of CMB photons with free electrons in the late-time Universe.  The \textit{thermal} SZ (tSZ) effect results when the electrons have high temperature, while the \textit{kinematic} SZ (kSZ) effect results from electrons with non-zero bulk velocity with respect to the CMB frame.  Cross-correlations of CMB data with measurements of late-time structure offer the prospect of measuring these two effects, and thereby accessing new information about the distribution, thermal state, and dynamics of baryons.  This information can in turn be used to improve cosmological constraints.

\vspace{0.1in}
\noindent \textbf{Constraining baryonic feedback} 

Although baryons make up only 16\% of the matter budget in the Universe, their impact on the matter distribution represents a challenge for cosmologists.  A combination of star formation, supernovae and ejecta from active galactic nuclei give rise to several complex and energetic process --- collectively known as \textit{feedback} --- that redistribute matter, causing significant impact (roughly 10\%) on the matter power spectrum at small scales.  Our inability to accurately model the impact of feedback  on the matter distribution limits our ability to use measurements on small scales to constrain cosmology \cite{DESy1:2017, Huang:2019, Amodeo:2020mmu, Amon2022}.  

The tSZ and kSZ effects offer the prospect of directly probing the diffuse ionized gas that is so impacted by feedback, and thus improving our understanding of its effects on the matter distribution.   The tSZ effect is sensitive to the electron gas pressure, and is therefore sensitive to changes in the thermal energy of the gas and its distribution.  The kSZ, on the other hand, offers the possibility of measuring the ionized gas density.  

Cross-correlations are essential to this program, since the tSZ and kSZ by themselves provide only a line-of-sight integral of the ionized gas properties.  By cross-correlating tSZ measurements with low-redshift structure, gas properties at different redshifts can be extracted, enabling powerful tests of feedback models \cite{Hill:2018,Pandey:2020,Pandey:2021,Troster:2021}.  The kSZ effect, on the other hand, depends on both the gas density and its velocity.  Only by using prior information on the gas velocity --- from, for example, galaxy surveys --- can information about the gas density be extracted.  A combination of kSZ and tSZ, together with mass distribution information from lensing, can be used to determine the full thermodynamic information of the halos, including the amount of feedback, the fraction of non-thermal pressure, and the temperature profile \cite{Battaglia:2017neq, Amodeo:2020mmu}. The direct access to the gas properties through the SZ effect allows for a direct calibration of baryon effects in weak lensing \cite{Amodeo:2020mmu, AtacamaCosmologyTelescope:2020wtv}. Moreover, ``projected fields'' techniques \cite{Ferraro:2016ymw, Hill:2016dta, Kusiak:2021hai} have been developed to handle photometric data and will reach their full potential with the next generation of cosmic surveys.

\vspace{0.1in}
\noindent \textbf{Cosmology with the kSZ effect} 

Since the kSZ signal is proportional to the galaxies' peculiar velocity, the latter can be reconstructed by using high resolution CMB maps together with galaxy catalogs \cite{Smith2018}. The reconstruction typically has lower noise on larger scales, often the most affected by primordial physics.
For example, the cross-correlations of the kinematic SZ effect with an overlapping galaxy survey can yield tight constraints on local primordial non-Gaussian fluctuations, characterized by the parameter $f_{\rm{NL}}$ \cite{Munchmeyer2018}. Reaching a target of $\sigma(f_{\rm{NL}}) < 1$ would disfavor a wide class of multi-field inflation models, shedding light on the nature of inflation~\cite{Alvarez:2014vva, Ferraro:2014jba,Smith2018,Munchmeyer2018,Deutsch:2017ybc,Contreras2019,Cayuso2018}.  Cross-correlations of galaxy surveys, such as LSST, with proposed CMB experiments, such as CMB-S4 and CMB-HD have the potential to reach this exciting threshold~\cite{Munchmeyer2018,Sehgal:2019nmk,CMB-HD-Snowmass,Snowmass2021:Inflation}.

\vspace{0.1in}
\noindent \textbf{Polarized SZ} 

The polarized SZ effect~\cite{Sazonov:1999zp} describes the process by which CMB polarization arises along the line of sight to galaxies and clusters due to the scattering by free electrons in the objects when the incident CMB intensity exhibits quadrupolar anisotropy.
Upcoming CMB surveys such as CMB-S4 and CMB-HD will have the sensitivity necessary to detect and characterize this effect~\cite{Hall:2014wna,Deutsch:2017cja,Louis:2017hoh,Meyers:2017rtf,S4,CMB-HD-Snowmass}.  The combination of future galaxy surveys and CMB surveys will allow for the reconstruction of three-dimensional maps of the remote temperature quadrupoles, thereby enabling several cosmological applications of the polarized SZ effect including: accessing additional information about cosmological fluctuations on the largest scales~\cite{Kamionkowski:1997na,Seto:2005de,Deutsch:2017ybc,Meyers:2017rtf}, probing late time structure formation and dark energy~\cite{Cooray:2002cb,Cooray:2003hd}, measuring the baryon content of galaxy clusters~\cite{Louis:2017hoh}, and searching for primordial gravitational waves~\cite{Alizadeh:2012vy,Deutsch:2018umo}.

\subsection{Cross-correlations with Line Intensity Mapping}

Line intensity mapping (LIM) is an emerging and potentially powerful observational technique to map the LSS over a wide range of scales and redshifts \cite{kovetz2017}. LIM detects the cumulative, unresolved emission of molecular and atomic spectral lines from galaxies together with the intergalactic medium. Measurements of the line frequency and spatial fluctuations in the line intensity provide a 3D map of the underlying dark matter distribution. Leveraging synergies between LIM at millimeter wavelengths and LIM of the neutral hydrogen 21cm line, together with optical galaxy surveys and CMB lensing, can significantly enhance the scientific return from each probe. The expected gain of cross-correlation analyses is the result of degeneracy breaking between cosmological parameters, sample variance cancellation, control of systematics, and improved calibration of nuisance parameters.

\vspace{0.1in}
\noindent \textbf{LIM $\mathbf{\times}$ galaxies} 

The cross-correlation between mm-wave LIM and upcoming optical galaxy surveys is a promising means by which to reduce systematic uncertainties on large scales, such as from Galactic dust extinction and stellar contamination. For mm-wave surveys that accurately recover low $k_\parallel$ modes, this cross-correlation can furthermore improve the calibration of the redshift distribution in imaging surveys by reducing uncertainties in photometric redshift measurements (i.e., ``clustering-based'' redshift estimation \cite{Menard:2013aaa,Alonso:2017dgh,Cunnington:2018zxg,Guandalin:2021sxw}). The potential strength of LIM for this task is twofold: accurate measurement of redshifts together with wide redshift coverage extending well beyond the spectroscopic surveys.

\vspace{0.1in}
\noindent \textbf{LIM $\mathbf{\times}$ CMB lensing} 

While CMB lensing is an observationally clean signal, the information content is limited due to the breadth of the projection kernel along the line of sight, with significant weighting of modes beyond $z\sim 2.$ LIM cross-correlations provide a unique opportunity to derive tomographic information at very high redshift that is not achievable with most spectroscopic galaxy surveys; unlocking this potential will require a dedicated effort to clean large-scale modes along the line of sight from foreground contamination via a variety of reconstruction methods \cite{Zhu:2015zlh,Zhu:2016esh,Modi:2019hnu,Li:2020uug,Darwish:2020prn}. Such an effort would be quite fruitful: by mitigating degeneracies between growth, line bias and mean brightness temperature, and through exploitation of  cosmic variance cancellation, these cross-correlations can significantly improve the constraints on early dark energy, modified gravity, neutrino mass measurements, and local non-Gaussianity  \cite{Schmittfull:2017ffw,Yu:2018tem,Wilson:2019brt,Sailer2021}.

\vspace{0.1in}
\noindent \textbf{LIM $\mathbf{\times}$ LIM} 

Although 21cm auto-spectra will face challenges from Galactic foreground contamination, cross-correlating intensity maps between 21cm and other lines (such as CO and/or [CII]) is a potentially powerful spectroscopic tracer at $z>3$ that could provide convincing evidence of the cosmological origin of high-redshift 21cm emission \cite{Lidz:2011dx}. Multi-line cross-correlations also open up the possibility of marginalizing over astrophysical uncertainty associated with the Epoch of Reionization (e.g., by tracing the scale at which the cross-correlation changes sign) \cite{Chang:2019xgc, Gong:2011mf}. Furthermore, considering higher-order cross-statistics can potentially improve the reliability with which line bias factors can be extracted, while presenting challenges that are interesting in their own right \cite{Beane:2018pmx}.

\subsection{Cross-correlations between imaging and spectroscopic surveys}

Imaging and spectroscopic surveys provide highly complementary information which, when combined over the same area of sky, can significantly improve upon the capabilities of either alone.  Joint-analyses of overlapping imaging and spectroscopic survey datasets will allow new tests of cosmology and fundamental physics, and can make core cosmological studies more robust to systematic uncertainties.

Below, we summarize some of the ways that 
combining data from a wide-field, highly-multiplexed optical and near-infrared multi-object spectroscopic (MOS) survey (e.g. DESI, 4MOST, MegaMapper, MSE, Roman) and from overlapping photometric surveys (e.g. LSST, Roman, Euclid) can significantly increase the return from both datasets and unlock additional scientific opportunities. 

\vspace{0.1in}
\noindent \textbf{Photometric redshift calibration} 

Photometric redshifts are a critical tool for imaging surveys. If these redshift estimates have an undetected systematic bias, dark energy inference can be catastrophically affected \citep[see e.g.,][]{Hearin2010,LSSTDESC2018}. Direct calibration via a large, representative spectroscopic redshift sample may not be possible given the depth of future imaging surveys \citep{Newman2015}. However, methods based upon cross-correlating the locations of photometric galaxies with spectroscopic samples can provide an alternative route for photometric redshift calibration \citep{Newman2008}.

\vspace{0.1in}
\noindent \textbf{Characterizing intrinsic alignments}

Intrinsic alignments of galaxy shapes
are a known contaminant to weak gravitational lensing \citep{Brown2002,Troxel2015}. If not accounted for, their presence can generate significant biases in cosmological analyses \citep{Kirk12, Krause16, Blazek:2019,Yao2017}. With a wide-field MOS we can measure the cross-correlation between positions of bright galaxies with spectroscopy and intrinsic shapes of fainter galaxies used for lensing, constraining IA models \citep{2014MNRAS.445..726C,Singh14,Johnston18}.

\vspace{0.1in}
\noindent \textbf{Characterizing strong lensing systems}

Upcoming imaging surveys will discover significantly more strong gravitational lenses than are currently known \citep[e.g.,][]{Collett15}. Strong lensing science requires redshifts for both lens and source. Many lenses are bright enough for redshift measurements via targeted fibers within very wide-area surveys, enabling identification of the systems best suited for follow-up observations as described in \citep{single}.

\vspace{0.1in}
\noindent \textbf{Spectroscopy for supernova cosmology}

Type Ia supernovae (SNe Ia) provide a mature probe of the accelerating universe \citep[e.g.,][]{2018arXiv181102374D}, and their use as standardizable candles is an immediate route to measuring the equation of state of dark energy. However, a major systematic uncertainty is the photometric classification and redshift measurement of the supernovae. Wide-field spectroscopy can address this in two ways.  First, spectroscopic observations are used for the classification of live SNe and the construction of optimized, large, homogeneous and representative training sets needed for purely photometric classifiers that may be used for the next generation of SN Ia cosmology \citep[e.g.,][]{2016ApJS..225...31L}.  Second, spectroscopy is used to obtain redshifts for host galaxies of SNe that have faded away.  While conventional SN Ia cosmology analyses rely on spectroscopic follow-up of live SNe, new analyses \citep[e.g.][]{jones2018,campbell2013, hlozek2012} show that it is possible to take advantage of even larger samples of SNe after obtaining spectroscopic redshifts of their host galaxies.

\vspace{0.1in}
\noindent \textbf{Testing General Relativity on cosmological scales} 

Combining cross-correlations between galaxy density and lensing with measurements of redshift-space distortions in galaxy clustering allows for tests of gravity on cosmological scales \citep{Ishak2019}. A sample of weak lensing galaxies with spectroscopic overlap would enable measurement of statistics sensitive to the nature of gravity, such as $E_{G}$ 
\cite{Zhang2007,Reyes2010}. For example, combining LSST and DESI/4MOST would enable multiple determinations of $E_G$ to $\sim 0.004$, roughly 10 times more precise than current constraints. 

\subsection{Multi-wavelength studies of galaxy clusters}
\label{sec:clusters}

Clusters of galaxies---the largest gravitationally-bound systems in the universe---are prominent tracers of cosmic structure.   
The abundance, spatial distribution, and other properties of these massive systems are highly sensitive to the physical laws and phenomena that govern how structure grows over time. 
Clusters provide sensitive constraints on dark energy parameters, the sum of the masses of the neutrino species, primordial non-Gaussianity, and the density of dark matter (see e.g., \cite{allen11} for a recent review).  For some cosmological models, clusters have the potential to be the \textit{most} constraining probe \cite{2016arXiv160407626D}.  For these reasons, the ability to detect and characterize galaxy clusters is a key science driver of current and future cosmological surveys in the optical, submillimeter and $X$-ray.

There are several upcoming surveys that will provide cosmologically interesting cluster observations, each offering distinct advantages, and we highlight a few of the largest here as examples of the potential for cross-wavelength analyses. LSST and Euclid will provide deep observations of the sky at optical through infrared wavelengths. In addition to compiling large ($>100,000$) samples of clusters identified via galaxy overdensities, these surveys will provide critical weak lensing mass calibration and redshift information for cluster samples selected in other surveys. eROSITA and Athena represent the next generation of X-ray surveys. The ongoing all sky eROSITA mission is expected to  discover $\sim$100,000 galaxy clusters (primarily at lower redshifts) in a way that is highly complimentary to the optical and tSZ surveys. Athena's (scheduled for launch in 2031) unprecedented sensitivity will allow detection of galaxy clusters and groups to $z\sim 2$ in a modest area survey as well as extensive targeted characterization of clusters detected at other wavelengths \cite{2013arXiv1306.2307N}, with observables highly complimentary to those at other wavelengths.

\vspace{0.1in}
\noindent \textbf{Mitigation of systematic effects through multi-wavelength observations}

The main challenge confronting cluster cosmology is the presence of difficult to characterize systematic uncertainties.
The next decade offers an unprecedented confluence of  expansive new multi-wavelength cosmic surveys and new computational and simulation capabilities which, if fully leveraged, will improve control of these errors and enable us to maximize the potential of these extreme systems to probe the composition and physical laws of the universe.
As clusters are multi-component systems and traced by numerous signatures (e.g., from overdensities of galaxies at optical and infrared wavelengths, hot gas detectable at X-ray and millimeter wavelengths, and as significant mass peaks in lensing data), employing data from all available telescopes will enable analyses that are not susceptible to any single set of astrophysical systematics or observational biases.  For instance, optical imaging can provide precise weak lensing mass calibration of SZ-selected clusters \cite{Miyatake:2019,Stern:2019}.  SZ observations, on the other hand, can be used to improve the robustness of optical cluster selection \cite{Costanzi:2021,Grandis:2021}.  

\vspace{0.1in}
\noindent \textbf{High redshift clusters}

The nature of the inverse Compton scattering process that gives rise to the tSZ makes the effect independent of redshift, providing a means to detect galaxy clusters out to high redshift.
With their deep and wide fields covering a large amount of volume, and
the ultra-deep fields imaging lower-mass clusters, CMB-S4 and CMB-HD will provide effective probes of the crucial regime of $z \gtrsim 2$, when galaxy clusters were vigorously accreting new hot gas while at the same time forming the bulk of their stars.  The CMB-S4 and CMB-HD catalogs will be more than an order of magnitude larger than current catalogs based on tSZ or X-ray measurements, and will contain an order of magnitude more clusters at $z > 2$ than will be discovered with current surveys.  Additionally, the gravitational lensing maps reconstructed from CMB survey data will provide unique mass calibration measurements for the highest redshift systems.

\vspace{0.1in}
\noindent \textbf{Combining with spectroscopic surveys}

Finally, while not expected to be drivers of cluster catalog production, wide-area spectroscopic surveys such as DESI and future more ambitious surveys will also play critical roles in mitigating systematic biases in next generation cluster analyses as large spectroscopic samples are crucial for calibrating photometric redshifts for both the cluster samples themselves as well as those of the source galaxies used in weak lensing mass calibration.  Spectroscopic surveys are also important for constraining certain cluster cosmological systematics (e.g., quantifying the contamination to optical cluster samples from line-of-site structure).

\section{Modeling \& Analysis challenges}
\label{sec:model_analysis}

Through measurements made with a panoply of instruments, cosmological data in the next decade will usher in a new era of large-scale, multi-wavelength, and overlapping surveys that opens up the exciting prospect of \emph{analyzing all datasets simultaneously}. As we have demonstrated above, significant additional information can be extracted from correlations between these datasets.  However, modeling multi-survey correlations necessarily requires additional work beyond that typically undertaken by single surveys.  Here we highlight some of the unique challenges presented by such analyses.

Part of the power of cross-survey analyses will come from the fact that they utilize information from multiple surveys at once from both linear and non-linear scales. Theoretical forecasts of the cosmological constraining power of the nonlinear regime now date back many years \cite{zentner_etal13,reid2014,krause_etal17,Shirasaki2020,salcedo2020}, and indicate that using smaller-scale information can result in factors of 2-4 improvement on dark energy constraints beyond present-day capabilities. Of course, the natural question arises as to whether these gains can be realized in practice, or whether the need to marginalize over nuisance parameters capturing systematic uncertainty result in an excessive loss of constraining power. Recent work analyzing BOSS galaxy samples has shown that these gains could indeed be a reality \cite{wibking_etal20,lange_hearin_2021,chapman_etal21}; by including measurements from nonlinear scales, these recent analyses have achieved a full factor of 2 improvement beyond previous BOSS analyses that restricted attention to the quasi-linear regime. Harvesting information from nonlinear scales nonetheless requires an expansion of the model parameter space, and so incorporating multi-wavelength capabilities into these models would enable analyses to leverage information from multiple surveys to break nuisance parameter degeneracies.  There is a clear opportunity for synergy in the development of these capabilities, since numerous forecasts have also established the potential to achieve comparable gains by leveraging multi-wavelength large-scale structure measurements that jointly analyze galaxy clusters \cite{salcedo_etal20_cluster_crossx,nicola_etal20,eifler_etal21}.

Further enhancements beyond standard analyses of large-scale structure come from utilizing higher-order statistics. While second-order statistics such as clustering and lensing have been the default method in analyzing cosmological data to date, higher-order statistics are expected to unveil new information about astrophysics \cite{behroozi_etal21}, the galaxy--halo connection \cite{tinker_etal08,wang_etal19}, as well as cosmology \cite{Uhlemann_etal20,banerjee_abel_2020}. Carrying out such analyses together with multi-survey cross-correlations has potential to improve control over systematic effects such as fiber collisions \cite{guo_etal12_fiber_collisions} and baryonic effects \cite{foreman_etal20}, and so in order to achieve maximal and robust returns from higher-order statistics, it will be necessary for the community to invest at qualitatively new levels in the development of sophisticated modeling efforts with capability to address these challenges.

Contemporary efforts to derive cosmological constraints from nonlinear scales (both with and without use of higher-order statistics) are typically built upon simplistic empirical models such as the Halo Occupation Distribution (HOD). Due to the very formulation of HOD-type models, incorporating new constraints from more than a single tracer galaxy population requires a significant expansion of the parameter space, and/or reliance upon plausibly-violated assumptions about the galaxy-halo connection. Thus conventional halo occupation models actually {\em penalize} attempts to incorporate new constraining data. This older generation of models was devised at a time when the reliability of cosmological simulations to resolve halo substructure was not yet established, and so halo occupation models are founded upon host halos identified at a particular simulated snapshot; considerable progress has been made during the intervening years on the quality of both cosmological simulations as well as the associated data products, and subhalo catalogs with merger trees are becoming widely available for high-resolution, survey-scale simulations \cite{Chuang_etal19_unit_sims,heitmann_etal_last_journey,Ishiyama_etal21_uchuu,bose_etal21}. In this sense, conventional techniques designed to harvest cosmological information in the nonlinear regime such halo occupation models bear the mark of the single-survey era in which they were developed.

Historically, generating physically realistic multi-wavelength predictions has required modeling approaches such as {\em hydrodynamical simulations} or {\em Semi-Analytic Models} (SAMs). While such models remain irreplaceable in the effort to understand the detailed physics of galaxies and clusters, the scientific payload of multi-wavelength cross-correlations can only be delivered with expansive explorations of parameter space based on high-resolution, Gpc-scale simulations, and so direct constraints based on traditional implementations of these models may be out of reach for the 2020s. Thus,  \textit{theoretical techniques with practical capability to conduct complete, multi-survey cosmological inference currently do not exist, and so the field of theoretical cosmology is ill-equipped for the quality, richness, and volume of cosmological data that will arrive in the 2020s.} 

Considerable recent progress has been made by a new generation of empirical models that bridge the gap between the level of complexity achieved by SAMs and the computational efficiency of empirical models, e.g., UniverseMachine \cite{behroozi_etal18} and EMERGE \cite{moster_etal17}. The ability of these models to make CPU-efficient predictions across redshift is quite promising, but significant further advances are needed on both the modeling and computation side for this new approach to conduct multi-wavelength inference with survey-scale simulations and emerging computing architectures. Towards this end, an emerging trend spanning numerous fields in computational science \cite{Kochkov2021_ML_CFD,jaxmd2020,hafner_veros_2018} is to build prediction pipelines within software frameworks for automatic differentiation such as JAX \cite{jax2018github} and TensorFlow \cite{tensorflow2015_whitepaper}, which are being actively developed to support the performance needs of contemporary deep learning applications. This approach to generating  autodiff-based predictions has now been applied in a variety of cosmological applications, including simulations of the density field \cite{modi_lanusse_seljak_2021_flowpm}, halo models \cite{jax_cosmo}, and simulation-based modeling of galaxy SEDs \cite{hearin_etal21_dsps,hearin_etal21_shamnet}; in addition to creating the capability to leverage gradient information via autodiff, large-scale structure pipelines constructed in this fashion naturally leverage the performance of these libraries on GPUs and other accelerator devices, and thereby anticipate the computing resources that will be available in the 2020s. In order to meet the predictive needs associated with the incoming flood of multi-wavelength astronomical data, and to maximize the scientific returns of the upcoming surveys, we consider it critical and urgent for the cosmology community to invest in the development of a new generation of modeling approaches that builds upon this progress and addresses the key limitations of contemporary techniques.

Beyond the technical challenges associated with cross-survey analyses, there are also practical difficulties associated with this work.  Any such analysis necessarily requires detailed knowledge of data products generated by multiple surveys.  Some of this information may be proprietary, and not easily shared.  Previous cross-survey analyses have typically waited until data products become public (thereby delaying results) or have operated through cross-survey memoranda of understanding (MoU).  Relative to single-survey analyses, analyses conducted through MoU are often subject to additional bureaucratic hurdles that can delay progress and unnecessarily increase workloads.  These difficulties can be significant enough to discourage cross-survey analyses, a clearly suboptimal outcome.

\section{Growth Opportunities}
\label{sec:recs}

In the previous sections we have detailed a wide range of opportunities enabled by combining and cross-correlating large, multi-wavelength, datasets from multiple cosmological experiments. In particular, joint-probe analyses across surveys have the potential to provide complementary information to single-probe experiments that could otherwise be limited by astrophysical and observational systematic effects.  We have also highlighted some of the challenges that such analyses face. To capitalize upon these opportunities and address the associated challenges, a qualitatively new level of investment in cross-survey, joint-probe infrastructure is required -- this includes simulations, associated modeling, coordination of data sharing, survey strategy, and training for the next-generation of scientists in a way that transcends any individual project or collaboration. The required investments are substantial, but they are critical for the next generation of cosmic surveys to fully realize their potential. Below we present a summary of future opportunities for growth that have potential to multiplicatively enhance the scientific returns of cosmological surveys in the 2020s:

\begin{itemize}
    \item \textbf {Joint simulations:} Nearly all of the multi-probe analyses discussed above require high-fidelity synthetic data that is validated against observational data. The computational demands of these simulations can be extremely expensive, and an intensive human resource effort is required in order to generate synthetic data that is sufficiently high-quality to merit this expense. Considerable progress has been made in this area in recent years, but efforts are typically limited to an individual survey, or even an individual probe in isolation. For example, most CMB simulations do not include physically realistic models of galaxy populations at low redshift, and synthetic datasets tailored for optical surveys of galaxies do not commonly include realistic treatments of the diffuse gas that can be observed in CMB surveys via, e.g., the SZ effect.
    As a result, there is a steeply increasing need in the field for simulations that are suitable for multi-wavelength cross-correlation analyses. This widespread need reflects a key opportunity for further growth in the area of generating multi-survey synthetic data, and the wider cosmology community stands to greatly benefit from substantially increased support for these efforts.
    \item \textbf{Joint modeling and analysis:} Current toolkits such as \texttt{Cobaya} \cite{Cobaya}, \texttt{Monte Python} \cite{MontePython}, \texttt{CosmoLike} \cite{cosmolike}, and \texttt{CosmoSIS} \cite{Zuntz:2015} have been successful in combining a number of ``standard'' large-scale structure probes and deriving posteriors on cosmological parameters through Bayesian analyses. Sophisticated modeling efforts with capability to make multi-wavelength predictions that leverage high-resolution simulations are commonly implemented in custom codebases that require highly specialized techniques in order to infer cosmological parameters in a Bayesian fashion. Here there exists another exciting opportunity to fully integrate a new generation of simulation-based models together with cosmological inference pipelines, leveraging new technologies such as machine learning methods, GPU interfaces, automatic gradient approaches, and likelihood-free inference methods. 
    \item \textbf{New initiatives enabling joint analyses:} By construction, multi-survey analyses in the era of large collaborations are often not hosted under one single collaboration with well-established communication structure and analysis tools. In the present day, such analyses are often enabled by MoUs and other agreements, or carried out with public data. This structure could create an inherent barrier for multi-survey analyses, and suppress potential opportunities for exciting discoveries, while new levels of effort in cross-survey collaboration could offer major benefits to the scientific returns of future surveys. Such initiatives could include coordination of survey strategy to ensure overlap, joint-processing of data, and coordination of cross-survey blinding strategies. New funding lines that focus on multi-survey cross-correlation analyses could be an effective, modest way to address some of these limitations.  The scope of these problems, however, warrants consideration of new ``centers'' focusing on development of joint simulation/modeling/analysis tools, as well as training/education for the next generation cosmologists who will be confronted with data in the 2020s that is of a qualitatively new character from previous decades. 
    \item \textbf{Support for proposed cosmic survey instruments:} The enormous potential of joint analyses discussed in this white paper is necessarily built on the success of single-probe experiments.  Enabling cross-survey analyses requires support for wide-field cosmic surveys including those listed in Table~\ref{tab: survey_summary}, and many more described in accompanying Snowmass white papers \citep{CMB-HD-Snowmass,CF4_DESI2_Snowmass,CF3_DM_facility_Snowmass, mmLIM_Snowmass}.   In return, joint-probe analyses will provide critical and complementary information to the understanding of cosmic acceleration and other fundamental physics. 
\end{itemize}

\bibliographystyle{JHEP}
\bibliography{thebib.bib}

\end{document}

%% file: authors.tex

\author{Eric J. Baxter}
\affiliation{\footnotesize Institute for Astronomy, University of \hawaii, Honolulu, HI 96822, USA}

\author{Chihway Chang}
\affiliation{\footnotesize Department of Astronomy and Astrophysics, University of Chicago, Chicago, IL 60637, USA}
\affiliation{\footnotesize Kavli Institute for Cosmological Physics, University of Chicago, Chicago, IL 60637, USA}

\author{Andrew Hearin}
\affiliation{\footnotesize High Energy Physics Division, Argonne National Laboratory, 9700 South Cass Avenue, Lemont, IL 60439, USA}

\author{Jonathan Blazek}
\affiliation{\footnotesize Department of Physics, Northeastern University, Boston, MA 02115, USA}

\author{Lindsey E. Bleem} 
\affiliation{\footnotesize High Energy Physics Division, Argonne National Laboratory, 9700 South Cass Avenue, Lemont, IL 60439, USA}
\affiliation{\footnotesize Kavli Institute for Cosmological Physics, University of Chicago, Chicago, IL 60637, USA}

\author{Simone Ferraro}
\affiliation{\footnotesize Lawrence Berkeley National Laboratory, One Cyclotron Road, Berkeley, CA 94720, USA}

\author{Mustapha Ishak}
\affiliation{\footnotesize Department of Physics, The University of Texas at Dallas, Richardson, TX 75080, USA}

\author{Kirit S. Karkare}
\affiliation{\footnotesize Kavli Institute for Cosmological Physics, University of Chicago, Chicago, IL 60637, USA}
\affiliation{\footnotesize Fermi National Accelerator Laboratory, Batavia, IL 60510, USA}

\author{Alexie Leauthaud}
\affiliation{\footnotesize Department of Astronomy and Astrophysics, University of California, Santa Cruz, 1156 High Street, Santa Cruz, CA 95064 USA}

\author{Jia Liu}
\affiliation{\footnotesize Kavli IPMU (WPI), UTIAS, The University of Tokyo, Kashiwa, Chiba 277-8583, Japan}

\author{Rachel Mandelbaum}
\affiliation{\footnotesize McWilliams Center for Cosmology, Department of Physics, Carnegie Mellon University, Pittsburgh, PA 15213, USA}

\author{Joel Meyers}
\affiliation{\footnotesize Department of Physics, Southern Methodist University, Dallas, TX 75275, USA}

\author{Azadeh Moradinezhad Dizgah}
\affiliation{\footnotesize D\'epartement de Physique Th\'eorique, Universit\'e de Gen\`eve, 24 quai Ernest Ansermet, 1211 Gen\`eva 4, Switzerland}

\author{Daisuke Nagai} 
\affiliation{\footnotesize Department of Physics, Yale University, New Haven, CT 06520, USA} 

\author{Jeffrey A. Newman}
\affiliation{\footnotesize Department of Physics and Astronony and PITT PACC, University of Pittsburgh, Pittsburgh, PA 15260, USA}

\author{Yuuki Omori}
\affiliation{\footnotesize Kavli Institute for Cosmological Physics, University of Chicago, Chicago, IL 60637, USA}

\author{Neelima Sehgal}
\affiliation{\footnotesize Physics and Astronomy Department, Stony Brook University, Stony Brook, NY 11794, USA}

\author{Martin White}
\affiliation{\footnotesize Department of Physics, University of California, Berkeley, CA 94720, USA}

\author{Joe Zuntz}
\affiliation{\footnotesize Institute for Astronomy, University of Edinburgh, Edinburgh EH9 3HJ, UK}


\author{Marcelo A. Alvarez}
\affiliation{\footnotesize Lawrence Berkeley National Laboratory, One Cyclotron Road, Berkeley, CA 94720, USA}

\author{Camille Avestruz}
\affiliation{\footnotesize Department of Physics, University of Michigan, Ann Arbor, MI, 48109, USA}

\author{Federico Bianchini}
\affiliation{\footnotesize Kavli Institute for Particle Astrophysics and Cosmology, Stanford University}
\affiliation{\footnotesize SLAC National Accelerator Laboratory, Menlo Park, CA 94025, USA}

\author{Sebastian Bocquet}
\affiliation{\footnotesize University Observatory, Faculty of Physics, Ludwig-Maximilians-Universität, Scheinerstr. 1, 81679 Munich, Germany}

\author{Boris Bolliet}
\affiliation{\footnotesize Department of Physics, Columbia University, New York, NY 10027, USA}

\author{John E. Carlstrom}
\affiliation{\footnotesize Department of Physics, University of Chicago, 5640 South Ellis Avenue, Chicago, IL, 60637, USA}
\affiliation{\footnotesize Kavli Institute for Cosmological Physics, University of Chicago, Chicago, IL 60637, USA}
\affiliation{\footnotesize High Energy Physics Division, Argonne National Laboratory, 9700 South Cass Avenue, Lemont, IL 60439, USA}
\affiliation{\footnotesize Department of Astronomy and Astrophysics, University of Chicago, Chicago, IL 60637, USA}

\author{Cyrille Doux}
\affiliation{\footnotesize Université Grenoble Alpes, CNRS, LPSC-IN2P3, 38000 Grenoble, France}

\author{Alexander van Engelen}
\affiliation{\footnotesize School of Earth and Space Exploration, Arizona State University, Tempe, AZ 85287, USA}

\author{Tze Goh}
\affiliation{\footnotesize Institute for Astronomy, University of \hawaii, Honolulu, HI 96822, USA}

\author{Sebastian Grandis}
\affiliation{\footnotesize University Observatory, Faculty of Physics, Ludwig-Maximilians-Universität, Scheinerstr. 1, 81679 Munich, Germany}

\author{J. Colin Hill}
\affiliation{\footnotesize Department of Physics, Columbia University, New York, NY 10027, USA}

\author{Anja von der Linden}
\affiliation{\footnotesize Department of Physics and Astronomy, Stony Brook University, Stony Brook University, Stony Brook, NY 11794, USA}

\author{Abhishek S. Maniyar}
\affiliation{\footnotesize Center for Cosmology and Particle Physics, Department of Physics, New York University, New York, NY 10003, USA}

\author{Gabriela A. Marques}
\affiliation{\footnotesize Department of Physics, Florida State University, Tallahassee, Florida 32306, USA}

\author{Anna Porredon}
\affiliation{\footnotesize Center for Cosmology and Astro-Particle Physics, The Ohio State University, Columbus, OH 43210, USA}

\author{Judit Prat}
\affiliation{\footnotesize Department of Astronomy and Astrophysics, University of Chicago, Chicago, IL 60637, USA}
\affiliation{\footnotesize Kavli Institute for Cosmological Physics, University of Chicago, Chicago, IL 60637, USA}

\author{Naomi Robertson}
\affiliation{\footnotesize Institute of Astronomy, University of Cambridge, Madingley Road, Cambridge, CB3 0HA, UK}

\author{Emmanuel Schaan}
\affiliation{\footnotesize Lawrence Berkeley National Laboratory, One Cyclotron Road, Berkeley, CA 94720, USA}

\author{Shabbir Shaikh}
\affiliation{\footnotesize School of Earth and Space Exploration, Arizona State University, Tempe, AZ 85287, USA }

\author{Tae-hyeon Shin}
\affiliation{\footnotesize Department of Physics and Astronomy, Stony Brook University, Stony Brook University, Stony Brook, NY 11794, USA}

\author{Yuanyuan Zhang}
\affiliation{\footnotesize Mitchell Institute for Fundamental Physics and Astronomy, Texas A\&M University, 576 University Dr, College Station, TX 77845, USA}